\documentclass[conference]{IEEEtran}
\IEEEoverridecommandlockouts
\usepackage{cite}
\usepackage{amsmath,amssymb,amsfonts}
\usepackage{algorithmic}
\usepackage{graphicx}
\usepackage{textcomp}
\usepackage{xcolor}
\usepackage{adjustbox}
\usepackage{multirow}
\usepackage{array, booktabs}
\usepackage{balance}
\usepackage{hyperref}
\usepackage{makecell}

\usepackage{dsfont}
\allowdisplaybreaks

\def\BibTeX{{\rm B\kern-.05em{\sc i\kern-.025em b}\kern-.08em
    T\kern-.1667em\lower.7ex\hbox{E}\kern-.125emX}}

\newcommand{\etal}{\emph{et~al.}}
\DeclareMathOperator{\FL}{FL}
\DeclareMathOperator{\SIM}{sim}
\DeclareMathOperator{\MAX}{max}

\begin{document}

\title{Improving Pediatric Pneumonia Diagnosis with Adult Chest X-ray Images Utilizing Contrastive Learning and Embedding Similarity\\[-1.0ex]}

\author{
\IEEEauthorblockN{Mohammad~Zunaed$^\dagger$, Anwarul Hasan$^\ddagger$ and Taufiq~Hasan$^\dagger$$^*$}
\IEEEauthorblockA{Email: rafizunaed@gmail.com, ahasan@qu.edu.qa, and taufiq@bme.buet.ac.bd}
\IEEEauthorblockA{$^\dagger$mHealth Lab, Dept. of Biomedical Engineering, Bangladesh University of Engineering and Technology, Dhaka, Bangladesh.}
\IEEEauthorblockA{$^\ddagger$Department of Mechanical and Industrial Engineering, Qatar University, Doha, Qatar.}
\IEEEauthorblockA{$^*$Department of Biomedical Engineering, Johns Hopkins University, Baltimore, MD.}
\\[-3.0ex]
}


\maketitle

\begin{abstract}
Despite the advancement of deep learning-based computer-aided diagnosis (CAD) methods for pneumonia from adult chest x-ray (CXR) images, the performance of CAD methods applied to pediatric images remains suboptimal, mainly due to the lack of large-scale annotated pediatric imaging datasets. Establishing a proper framework to leverage existing adult large-scale CXR datasets can thus enhance pediatric pneumonia detection performance. In this paper, we propose a three-branch parallel path learning-based framework that utilizes both adult and pediatric datasets to improve the performance of deep learning models on pediatric test datasets. The paths are trained with pediatric only, adult only, and both types of CXRs, respectively. Our proposed framework utilizes the multi-positive contrastive loss to cluster the classwise embeddings and the embedding similarity loss among these three parallel paths to make the classwise embeddings as close as possible to reduce the effect of domain shift. Experimental evaluations on open-access adult and pediatric CXR datasets show that the proposed method achieves a superior AUROC score of 0.8464 compared to 0.8348 obtained using the conventional approach of join training on both datasets. The proposed approach thus paves the way for generalized CAD models that are effective for both adult and pediatric age groups.
\end{abstract}


\section{Introduction}
According to the World Health Organization (WHO), pneumonia is one of the single largest causes of child mortality across the world \cite{pne_stat_cite}. Chest radiography (CXR) is the most frequently used imaging modality for diagnosing disease in children due to its affordability and availability \cite{Padash2022}. Deep learning-based computer-aided (CAD) diagnosis systems have demonstrated remarkable performance in analyzing adult CXRs, thanks to the availability of large-scale, annotated datasets \cite{xnet, thoraXNet}. However, despite the success of the development of diagnostic models for thoracic diseases on adult CXRs, research into the application of CAD systems to pediatric imaging remains in its infancy, especially due to the lack of large-scale pediatric datasets \cite{MORCOS2023742, Pham2023}.\par
Over the recent years, a number of deep-learning-based methods have been proposed for pneumonia diagnosis in pediatric CXRs. Prakash \etal \cite{Prakash2023} utilized two-stage training, i.e., extracted features from the deep learning model, Xception, and passed them to kernel principal component analysis and a number of classical models and MLP classifiers for final prediction. Chen \etal \cite{Chen2020} utilized a classifier model based on a convolutional neural network and compared it with different schemes, i.e., the one-versus-one scheme and the one-versus-all scheme for diagnosis of common pulmonary diseases in children by CXR images. An extensive review of deep learning-based methods for pediatric image analysis can be found in \cite{Padash2022}. The previous methods are either based on the pediatric CXR dataset only or utilized joint training of pediatric and adult CXR datasets. However, Morcos \etal \cite{MORCOS2023742} demonstrated that while a model trained with adult CXRs can adequately diagnose pneumonia in pediatric patients, models trained exclusively on pediatric CXRs performed better. This is expected because there is a domain gap between pediatric and adult-based CXRs. As the unavailability of large-scale datasets for pediatric diseases is still a hindrance to the development of pediatric-focused artificial intelligence (AI), leveraging adult-based large-scale datasets can expedite pediatric AI research. However, joint training of pediatric and adult-based datasets may result in sub-optimal performance due to class imbalance and domain mismatch issues. To the best of our knowledge, at an architectural level, this issue of the adult vs pediatric CXR domain gap has not been addressed by previous researchers.\par
\begin{figure*}[!t]
	\centering
	\includegraphics[width=0.85\linewidth]{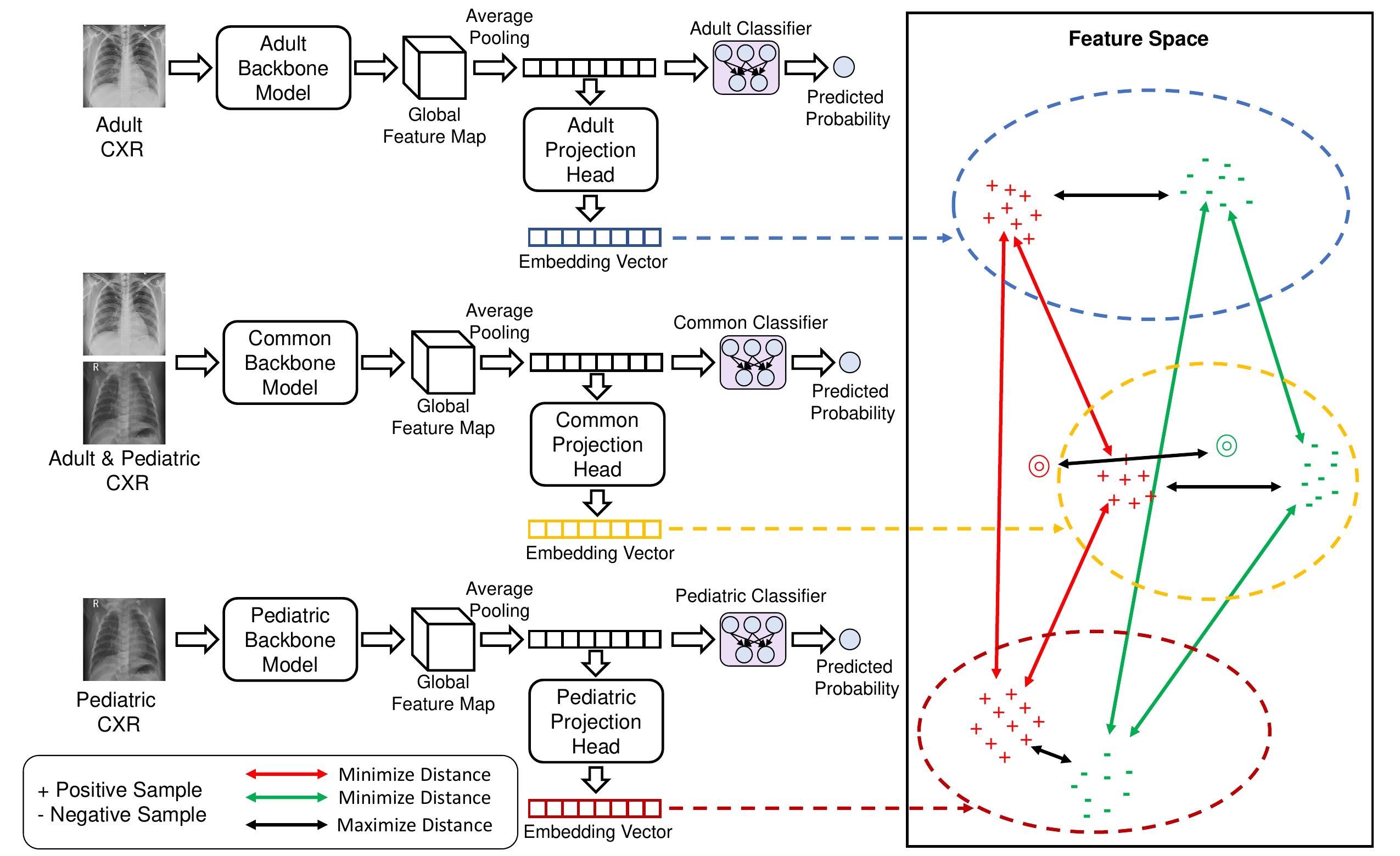}
        \caption{Overview of the proposed framework. The backbone models are based on ResNet-50 architecture. The adult and pediatric backbone models take adult and pediatric CXR images as input, respectively, while the common backbone model takes both adult and pediatric CXR images. Projection heads are used with the pooled global feature map to generate the embeddings. Three separate classifiers are utilized for predicting pathology probability and classification losses. Finally, the contrastive and embedding losses are utilized in embedding feature vector space to cluster the classwise embeddings, both intra- and inter-models, to reduce the impact of the domain gap.}
 	\label{fig: overall_framework}
\end{figure*}
In this paper, we introduce a framework with three parallel paths with contrastive learning and embedding similarity losses. The paths are trained with pediatric only, adult only, and with both types of CXRs, respectively. Our motivation is that the jointly-trained model, which is trained with adult and pediatric datasets, is susceptible to this domain gap as it is trained with both types of datasets. However, as the adult-only or pediatric-only-based models are trained with only CXRs of their respective domains, domain information becomes less integrated as since all data are from the same domain (adult/pediatric), and thus does not help in classification. As a result, to minimize the domain gap of the main model, the classwise embeddings among all three paths need to be as close as possible. We utilize the multi-positive contrastive loss \cite{tian2023stablerep} for clustering embeddings on each path. A projection head is applied after the global average pooling of the model to generate the embeddings. As a similarity metric, we use the cosine similarity loss between the classwise embeddings of these three paths. We simultaneously train all three paths together, whereas models on each path are based on ResNet50 architecture. Experimental validations are done using open-access pediatric and adult CXR datasets to evaluate the effectiveness of the proposed architecture compared to standard joint training on data from both domains.

\section{Methodology}
\subsection{Problem formulation}
We have two training sets of different domains, i.e., pediatric domain, $\mathcal{D}_{p}$ consisting of $N_p$ samples, $D_p = \{(x^{(i)}_{p}, y^{(i)}_{p});i=1,\ldots,N_p\}$ and adult domain $\mathcal{D}_{a}$ consisting of $N_a$ samples, $D_a = \{(x^{(i)}_{a}, y^{(i)}_{a});i=1,\ldots,N_a\}$. Here, each input CXR image $x^{(i)}$ is associated with a ground truth label $y^{(i)} \in [0, 1]$. In addition, we have a pediatric test dataset, $\mathcal{D}^{test}_{p}$, with $N^{test}_{p}$ samples. Our task is to learn a framework that will yield a deep learning model, $f^\phi(f^\theta(x)) \rightarrow y$, that can utilize the adult domain training set $\mathcal{D}_{a}$ with the pediatric domain training set $\mathcal{D}_{p}$ in order to improve performance on the $\mathcal{D}^{test}_{p}$ compared to a simple joint training method. Here, $f^\theta(\cdot)$ is the backbone model, and $f^\phi(\cdot)$ is the classifier.
\subsection{Overall framework}
The overall architecture of our proposed framework is illustrated in Fig. \ref{fig: overall_framework}. Our proposed framework consists of three paths: one main path with two auxiliary paths. Each path contains a backbone model and a classifier model. The main path takes both pediatric and adult CXR images as input, while the auxiliary paths take pediatric and adult CXR images as input, respectively. Each backbone model has a projection head attached to it to generate the embedding vector. The role of the auxiliary models is to help the main model reduce its bias toward domain information and focus on pathology markers more than disease markers. We hypothesize that as the pathology information is common among these parallel paths while there are input domain variations, the generated high-level disease-wise embeddings from each of the backbone models should be clustered together. We utilize contrastive learning to cluster the intra-model pathologies while use the embedding cosine similarity to cluster the inter-model pathologies.
\subsection{Contrastive Learning}
We utilize the multi-positive contrastive learning from the StableRep implementation \cite{tian2023stablerep}. Let's assume an anchor sample $z^a$, and a set of other sample candidates $\{z^{b_1}, z^{b_2}, \ldots, z^{b_N}\}$. We calculate the contrastive categorical distribution $q$ to find out to what extent the anchor sample $z^a$ matches each $z^b$ sample:
\begin{equation}
    q^{(i)} = \frac{\exp{(z^a \cdot z^{b_i}/\tau)}}{\sum_{j=1}^{N}\exp{(z^a \cdot z^{b_j}/\tau)}}
\end{equation}
where $\tau \in \mathcal{R}_{+}$ is the scalar temperature, and all the samples ($z^a$ and all $z^b$) are normalized by $l_2$. Afterward, we compute the ground-truth categorical distribution $p$, if the anchor sample is matched with at least one other sample, by:
\begin{equation}
    p^{(i)} = \frac{\mathds{1}_{match(z^a, z^{b_i})}}{\sum_{j=1}^{N}\mathds{1}_{match(z^a, z^{b_j})}}
\end{equation}
where the indicator function $\mathds{1}_{match(\cdot,\cdot)}$ indicates whether the anchor and candidate match. Intuitively, multi-positive contrastive learning loss is a $N$-way softmax classification distribution over all encoded sample candidates. Thus, the multi-positive contrastive loss is defined as the cross-entropy between the ground-truth distribution $p$ and the contrastive distribution $q$:
\begin{equation}
     H(p,q) = - \sum_{i=1}^{N} p^{(i)} \log q^{(i)}
\end{equation}
Our framework contains one common path with backbone model/feature extractor $f^\theta_c(\cdot)$ and the adult and pediatric path with feature extractor $f^\theta_a(\cdot)$ and $f^\theta_p(\cdot)$. We obtain the representations from the backbone models by,
\begin{align}
    h_c^{(i)} &= f^\theta_c(x_p^{(i)} \ or \ x_a^{(i)})\\
    h_p^{(i)} &= f^\theta_p(x_p^{(i)})\\
    h_a^{(i)} &= f^\theta_a(x_a^{(i)})
\end{align}
where $h^{(i)} \in \mathbb{R}^d$ is the output after the global average pooling layer. $d$ is the global dimension of the backbone model. Afterward, We add projection heads $(g_c(\cdot), g_p(\cdot), g_a(\cdot))$ that map these representations to the space where contrastive loss is applied. For the architecture of these projection heads, we adopt the small neural network projection head used in \cite{tian2023stablerep}. 
\begin{align}
    z_c^{(i)} &= g_c(h_c^{(i)})\\
    z_p^{(i)} &= g_p(h_p^{(i)})\\
    z_a^{(i)} &= g_a(h_a^{(i)})
\end{align}
This is a supervised learning setup where the labels of each CXR, i.e., whether they contain pathology or not, are known beforehand. We utilize these ground truth labels to generate the categorical distributions $p_c$, $p_p$, and $p_a$. We employ the projected representations/embedding vectors, $z_c$, $z_p$, and $z_a$ to generate contrastive categorical distributions, $q_c$, $q_p$, and $q_a$. Finally, the contrastive loss is formed by,
\begin{equation}
    \mathcal{L}_{cont} = H(p_c,q_c) + H(p_p,q_p) + H(p_a,q_a)
\end{equation}
\subsection{Classification Loss}
We utilize the focal loss, $\FL(\cdot,\cdot)$, as the classification loss \cite{8237586}. The representations, $h_c^{(i)}, h_p^{(i)},$ and $h_a^{(i)}$ are fed to the classifiers and sigmoid layer $\mathcal{S}(\cdot)$ to generate the probabilities. The classification loss is defined as,
\begin{align}
    \mathcal{L}_{cls} = & \FL(\mathcal{S}(f^\phi_c(h_c^{(i)})), y_c^{(i)})+\FL(\mathcal{S}(f^\phi_p(h_p^{(i)})), y_p^{(i)})+\nonumber\\
    &\FL(\mathcal{S}(f^\phi_a(h_a^{(i)})), y_a^{(i)})
\end{align}
Here, $ y_c^{(i)}, y_p^{(i)},$ and $y_a^{(i)}$ are the ground truths.
\subsection{Embedding Loss}
We take the average of the embeddings per class to generate the classwise embeddings $w$. Afterward, we calculate the embedding loss based on similarity and dissimilarity by,
\begin{align}
    \mathcal{L}_{emb}^{sim} &= \sum_{j=1}^{C}(2-\SIM(w^j_c, w^j_a) + \SIM(w^j_c, w^j_p))\\
    \mathcal{L}_{emb}^{dissim} &= \sum_{i=1,j=1}^{C}\mathds{1}_{[j \not\equiv i]}\MAX(0, \SIM(w^j_c, w^i_c))
\end{align}
Here, $C$ is the number of classes and $\SIM(\cdot,\cdot)$ denotes cosine similarity.

\section{Experiment and Result}
\subsection{Datasets \& Implementation details}
\subsubsection{Datasets}
We utilize the PediCXR dataset \cite{Pham2023} as the pediatric CXR dataset and the VinDr-CXR dataset \cite{nguyen2020vindrcxr} as the adult CXR dataset. The PediCXR dataset contains 9,125 CXR images, of which 481 are diagnosed with pneumonia pathology. The VinDr-CXR dataset contains 18,000 CXR images annotated by 17 experienced radiologists. To prepare the pneumonia label of the training split, we generate the positive labels based on the majority vote of the participating radiologists. Thus, the VinDr-CXR dataset contains 717 images diagnosed with pneumonia pathology. We utilize the official train and test split of these datasets provided by the authors. We split the training sets of both pediatric and adult CXR datasets into stratified 4-fold cross-validation schemes. 

\subsubsection{Implementation details}
CXR images often contain redundant information not pertinent to the pathology classification. As this extra information may impede the training, first, we train a U-Net-based lung segmentation model \cite{unet} using datasets from \cite{Shiraishi2000DevelopmentOA, VANGINNEKEN200619} to segment the lung regions \cite{Chen2020}. Next, we calculate the smallest bounding box that delimits both segmented lungs. We add 0.05\% pixels on all four sides of the bounding boxes based on the center coordinates. The CXR image is then cropped according to the resulting bounding box.\par
We utilize the transfer learning \cite{9607652} with ResNet50 as the backbone architecture pre-trained on the CheXpert dataset \cite{chexpert_ds}. We resize the CXR images to 224\texttimes224 and normalize them with the mean and standard deviation of the ImageNet training set \cite{imageNet}. We utilize horizontal flipping, random brightness, and contrast adjustment as augmentations. The models' parameters are updated using the AdamW optimizer \cite{adamW} with a weight decay of 0.0001 and a learning rate of 0.0001. The architecture is trained end to end for 50 epochs with a total batch size of 32 images, where the 1:1 ratio between pneumonia and non-pneumonia and 1:1 ratio between pediatric and adult CXR images are maintained in each iteration.
\subsubsection{Evaluation metrics}
We evaluate the classification performance by utilizing the area under the receiver operating characteristic curve (AUROC). The AUROC score reflects the degree of measure of separability, and the higher the AUROC achieves, the better the extent of separability.

\subsection{Experimental results \& quantitative analysis}
\subsubsection{Analyzing the Effect of Adult CXR Dataset}
First, we analyze the impact of the adult CXR dataset on the performance of the backbone model on the pediatric CXR test dataset. The results are given in Table \ref{table: Effect of the Adult CXR Dataset}. We can observe that while the adult CXR dataset alone can manage adequate performance on the pediatric test dataset, model training on the pediatric dataset achieves superior performance. This proves that there lies a domain gap between these two CXR datasets. Afterward, when utilizing the datasets together, we can see that the performance increases compared to single-domain training. While using only the adult CXRs does not result in superior performance, joint training with the pediatric CXRs improving the performance indicates that a proper method to use the adult CXR dataset may result in further performance improvements.

\begin{table}[!h]
\centering
\caption{Analysis of The Effect of the Adult CXRs on the Performance of the Backbone Model on the Pediatric Test Dataset.}
\label{table: Effect of the Adult CXR Dataset}
\begin{tabular}{| c | c | c |}
\hline
\multicolumn{2}{|c|}{Train Dataset} & \multirow{2}{*}{AUROC} \\
\cline{1-2}
Pediatric CXR & Adult CXR &  \\
\hline
\checkmark &   & 0.8211 \\
\hline
& \checkmark & 0.7972 \\
\hline
 \checkmark & \checkmark & 0.8348 \\
\hline
\end{tabular}
\end{table}

\subsubsection{Analyzing the Effect of Contrastive Loss}
Second, we add the contrastive loss to each model to evaluate the impact of the contrastive loss. The results are reported in Table \ref{table: Effect of the Contrastive Loss}. We can observe from the results that the contrastive loss improves the performance of all models, demonstrating the efficacy of clustering the embeddings. 

\begin{table}[!h]
\centering
\caption{Effect of the Contrastive Loss on the Performance of the Backbone Model on the Pediatric Test Dataset.}
\label{table: Effect of the Contrastive Loss}
\begin{tabular}{| c | c | c | c |}
\hline
\multicolumn{2}{|c|}{Train Dataset} & \multirow{2}{*}{\makecell{Contrastive \\ Loss}} & \multirow{2}{*}{AUROC} \\
\cline{1-2}
Pediatric CXR & Adult CXR &  &  \\
\hline
\checkmark &   &   & 0.8211 \\
\hline
\checkmark &   & \checkmark & 0.8349 \\
\hline
  & \checkmark &   & 0.7972 \\
\hline
  & \checkmark & \checkmark & 0.8053 \\
\hline
\checkmark & \checkmark &   & 0.8348 \\
\hline
\checkmark & \checkmark & \checkmark & 0.8381 \\
\hline
\end{tabular}
\end{table}

\subsubsection{Analyzing the Impact of Proposed Framework}
Finally, we utilize our proposed framework with three parallel paths utilizing both the contrastive and embedding losses. The results are reported in Table \ref{table: Effect of the Proposed Method}. We observe that our proposed framework improves the performance further, from 0.8381 to 0.8464, proving the effectiveness of the approach.

\begin{table}[!h]
\centering
\caption{Analysis of The Effect of the Proposed Method on the Pediatric Test Dataset with Joint Training Setup, Contrastive Loss, and Embedding Loss.}
\label{table: Effect of the Proposed Method}
\begin{tabular}{| c | c |}
\hline
Method & AUROC \\
\hline
Base &  0.8348 \\
\hline
Base + Contrastive Loss & 0.8381 \\
\hline
Base + Contrastive Loss + Embedding Loss & 0.8464 \\
\hline
\end{tabular}
\end{table}

\section{Conclusion}
In this paper, we have proposed a three-parallel path deep learning-based framework that can leverage the adult CXR datasets to improve the performance of the pediatric test datasets by utilizing the classwise embedding similarity between these parallel paths. Experimental results showed that our proposed framework could achieve 0.8464 AUROC compared to simple joint training of 0.8348.

\bibliographystyle{IEEEtran}
\bibliography{main.bib}

\begin{thebibliography}{10}
\providecommand{\url}[1]{#1}
\csname url@samestyle\endcsname
\providecommand{\newblock}{\relax}
\providecommand{\bibinfo}[2]{#2}
\providecommand{\BIBentrySTDinterwordspacing}{\spaceskip=0pt\relax}
\providecommand{\BIBentryALTinterwordstretchfactor}{4}
\providecommand{\BIBentryALTinterwordspacing}{\spaceskip=\fontdimen2\font plus
\BIBentryALTinterwordstretchfactor\fontdimen3\font minus \fontdimen4\font\relax}
\providecommand{\BIBforeignlanguage}[2]{{%
\expandafter\ifx\csname l@#1\endcsname\relax
\typeout{** WARNING: IEEEtran.bst: No hyphenation pattern has been}%
\typeout{** loaded for the language `#1'. Using the pattern for}%
\typeout{** the default language instead.}%
\else
\language=\csname l@#1\endcsname
\fi
#2}}
\providecommand{\BIBdecl}{\relax}
\BIBdecl

\bibitem{pne_stat_cite}
``World health organization: Pneumonia in children,'' \url{https://www.who.int/news-room/fact-sheets/detail/pneumonia/}, accessed: 02-02-2024.

\bibitem{Padash2022}
S.~Padash, M.~R. Mohebbian, S.~J. Adams, R.~D.~E. Henderson, and P.~Babyn, ``Pediatric chest radiograph interpretation: how far has artificial intelligence come? a systematic literature review,'' \emph{Pediatr Radiol}, vol.~52, no.~8, pp. 1568--1580, Jul 2022.

\bibitem{xnet}
U.~Kamal, M.~Zunaed, N.~B. Nizam, and T.~Hasan, ``Anatomy-{XN}et: An anatomy aware convolutional neural network for thoracic disease classification in chest {X}-rays,'' \emph{IEEE J Biomed Health Inform}, vol.~26, no.~11, pp. 5518--5528, 2022.

\bibitem{thoraXNet}
M.~I. Hossain, M.~Zunaed, M.~K. Ahmed, S.~M.~J. Hossain, A.~Hasan, and T.~Hasan, ``Thora{X}-{P}rior{N}et: A novel attention-based architecture using anatomical prior probability maps for thoracic disease classification,'' \emph{IEEE Access}, vol.~12, pp. 3256--3273, 2024.

\bibitem{MORCOS2023742}
G.~Morcos, P.~H. Yi, and J.~Jeudy, ``Applying artificial intelligence to pediatric chest imaging: Reliability of leveraging adult-based artificial intelligence models,'' \emph{J. Am. Coll. Radiol}, vol.~20, no.~8, pp. 742--747, 2023.

\bibitem{Pham2023}
H.~H. Pham \emph{et~al.}, ``Pedi{CXR}: An open, large-scale chest radiograph dataset for interpretation of common thoracic diseases in children,'' \emph{Sci. Data}, vol.~10, no.~1, p. 240, Apr 2023.

\bibitem{Prakash2023}
J.~A. Prakash, V.~Ravi, V.~Sowmya, and K.~P. Soman, ``Stacked ensemble learning based on deep convolutional neural networks for pediatric pneumonia diagnosis using chest x-ray images,'' \emph{Neural. Comput. Appl}, vol.~35, no.~11, pp. 8259--8279, Apr 2023.

\bibitem{Chen2020}
K.-C. Chen \emph{et~al.}, ``Diagnosis of common pulmonary diseases in children by x-ray images and deep learning,'' \emph{Sci. Rep}, vol.~10, no.~1, p. 17374, Oct 2020.

\bibitem{tian2023stablerep}
Y.~Tian, L.~Fan, P.~Isola, H.~Chang, and D.~Krishnan, ``Stablerep: Synthetic images from text-to-image models make strong visual representation learners,'' in \emph{Proc. Adv. Neural Inf. Process. Syst.}, vol.~36, 2023, pp. 48\,382--48\,402.

\bibitem{8237586}
T.-Y. Lin, P.~Goyal, R.~Girshick, K.~He, and P.~Dollár, ``Focal loss for dense object detection,'' in \emph{Proc. IEEE Int. Conf. Comput. Vis.}, 2017, pp. 2999--3007.

\bibitem{nguyen2020vindrcxr}
H.~Q. Nguyen \emph{et~al.}, ``Vin{D}r-{CXR}: An open dataset of chest {X}-rays with radiologist’s annotations,'' \emph{Sci. Data}, vol.~9, p. 429, 2022.

\bibitem{unet}
O.~Ronneberger, P.~Fischer, and T.~Brox, ``U-net: Convolutional networks for biomedical image segmentation,'' in \emph{Proc. Int. Conf. Med. Image Comput. Comput.-Assist. Interv.}, 2015, pp. 234--241.

\bibitem{Shiraishi2000DevelopmentOA}
J.~Shiraishi \emph{et~al.}, ``Development of a digital image database for chest radiographs with and without a lung nodule: receiver operating characteristic analysis of radiologists' detection of pulmonary nodules.'' \emph{AJR Am. J. Roentgenol.}, vol. 174(1), pp. 71--4, 2000.

\bibitem{VANGINNEKEN200619}
B.~{van Ginneken}, M.~B. Stegmann, and M.~Loog, ``Segmentation of anatomical structures in chest radiographs using supervised methods: a comparative study on a public database,'' \emph{Med. Image Anal.}, vol.~10, no.~1, pp. 19--40, 2006.

\bibitem{9607652}
T.~T. Tran \emph{et~al.}, ``Learning to automatically diagnose multiple diseases in pediatric chest radiographs using deep convolutional neural networks,'' in \emph{Proc. IEEE Int. Conf. Comput. Vis. Workshops}, 2021, pp. 3307--3316.

\bibitem{chexpert_ds}
J.~Irvin \emph{et~al.}, ``Che{X}pert: A {L}arge {C}hest {R}adiograph {D}ataset with {U}ncertainty {L}abels and {E}xpert {C}omparison,'' in \emph{Proc. AAAI Conf. Artif. Intell.}, 2019, pp. 590--597.

\bibitem{imageNet}
J.~Deng, W.~Dong, R.~Socher, L.-J. Li, K.~Li, and L.~Fei-Fei, ``Imagenet: A large-scale hierarchical image database,'' in \emph{Proc. IEEE Conf. Comput. Vis. Pattern Recognit.}, 2009, pp. 248--255.

\bibitem{adamW}
I.~Loshchilov and F.~Hutter, ``Decoupled weight decay regularization,'' in \emph{Proc. Int. Conf. Learn. Representations}, 2019.

\end{thebibliography}
\balance

\end{document}